%% file: Main.tex
\documentclass{IEEEcsmag}
\input{preamble}
\usepackage[colorlinks,urlcolor=blue,linkcolor=blue,citecolor=blue]{hyperref}
\usepackage{cleveref}
\expandafter\def\expandafter\UrlBreaks\expandafter{\UrlBreaks\do\/\do\*\do\-\do\~\do\'\do\"\do\-}
\usepackage[dvipsnames]{xcolor}
\usepackage[framemethod=tikz]{mdframed}

\begin{document}

\sptitle{Article Type: Description  (Special Issue on Engineering Agentic Systems)}

\title{Making Sense of AI Agents Hype: \\Adoption, Architectures, and Takeaways from Practitioners 

% \todo[inline]{1: Making Sense of the AI Agent Hype: What Practitioner Conferences Reveal About Real-World Architectures \\2: AI Agents in Practice: Adoption, Architectures, and Takeaways from Practitioner Conferences\\
% 3: Agent-based architectures in practice: lessons learned from practitioners
% \\4: Architecting AI-agents: lessons learned from practitioners
% Add authors affiliations}
}

\author{Ruoyu Su}
\affil{University of Oulu, Oulu, Finland}

\author{Matteo Esposito}
\affil{University of Oulu, Oulu, Finland}

\author{Roberta Capuano}
\affil{University of L’Aquila, L'Aquila, Italy}

\author{Rafiullah Omar}
\affil{University of L’Aquila, L'Aquila, Italy}

\author{June Sallou}
\affil{Wageningen University, Wageningen, The Netherlands}

\author{Henry Muccini}
\affil{University of L’Aquila, L'Aquila, Italy}

\author{Davide Taibi}
\affil{University of Southern Denmark, Vejle, Denmark \& University of Oulu, Oulu, Finland}

\markboth{THEME/FEATURE/DEPARTMENT}{THEME/FEATURE/DEPARTMENT}

\begin{abstract}
%AI agents have become one of the most prominent and rapidly adopted technologies in contemporary software engineering, driven by recent advances in large language models. However, their recent emergence leaves open questions about how organizations transition from experimentation to production, which architectural patterns guide agent-based design, and how such systems behave in real-world industrial practice. To support practitioners in navigating this fast-evolving topic, we present a review of practitioner conference talks on AI agents, analyzing companies’ adoption motivations, architectural choices, and their reported impact in industrial settings.

%To support practitioners in understanding how {\bf agentic systems} are designed in real-world industrial practice, we present a review of practitioner conference talks on AI agents. We analyze the transcript of 234 videos to report on how companies adopt agent-based architectures (Objective 1), capture ƒ√architectural knowledge for agentic systems (Objective 2), and assess the real-world impact and architectural implications of agentic systems (Objective 3).

%%%% V3 Rob %%%%
To support practitioners in understanding how {\bf agentic systems} are designed in real-world industrial practice, we present a review of practitioner conference talks on AI agents. We analyzed 138 recorded talks to examine how companies adopt agent-based architectures (\textit{Objective 1}), identify recurring architectural strategies and patterns (\textit{Objective 2}), and analyze application domains and technologies used to implement and operate LLM-driven agentic systems (\textit{Objective 3}).

%\todo[inline]{Davide: This is my first proposal. I don't like the wording of the Objective 3, I think we should find a better name for it, and then harmonize it in the whole paper. Matteo: done}

% \todo[inline]{SUBMISSION DEADLINE: 5 January 2026 }

%\todo[inline]{Note: this is a two-columns format (required for submission). The final version, instead, will be on a three-columns format. \\IEEE Software articles are typically structured into 5-8 sections, and eventually into smaller subsections. \\Rules from IEEE Software: Articles should be no more than 4,200 words, including 250 words for each figure and table. Inline equations will be counted with the overall word count. Blocks of code will count as 250 words. A maximum of 15 references and author biographies are not included in the word count. The abstract should be no more than 150 words and should describe the overall focus of your manuscript. With your submission, provide three actionable insights in bullet-list format that software practitioners will get from your paper. Please include a photo of each author. Note: Use American English when writing the paper.}

\end{abstract}

\maketitle

%old
%\chapteri{AI}
%  agents are rapidly reshaping the software engineering practices, supporting activities such as code generation, testing, documentation, workflow orchestration, and system operation.
%  \begin{ReviewerAEnv}The capabilities of AI agents are mainly due to recent advances in large language models (LLMs), which researchers continue to explore in depth their applications~\cite{franceschelli2025creativity} so that AI agents have become foundational components of modern AI-driven systems~\cite{naveed2025comprehensive,sajjadi2025survey}.\end{ReviewerAEnv}
% According to Gartner~\cite{gartner_ai_agents_2025}, 40\% of enterprise applications will feature task-specific AI agents by 2026, up from less than 5\% in 2025.
% \begin{ReviewerBEnv}International Data Corporation (IDC)\end{ReviewerBEnv}~\cite{IDC} also forecasts agent-based AI as a major driver of enterprise IT investment through large-scale agent orchestration.

% ROB Shrink
\chapteri{AI} agents are rapidly reshaping software engineering practices, supporting activities such as code generation, testing, documentation, workflow orchestration, and system operation.
\begin{ReviewerAEnv}These capabilities are largely driven by recent advances in large language models (LLMs), whose applications continue to be actively explored~\cite{franceschelli2025creativity}, making AI agents foundational components of modern AI-driven systems~\cite{naveed2025comprehensive,sajjadi2025survey}.\end{ReviewerAEnv}
According to Gartner~\cite{gartner_ai_agents_2025}, 40\% of enterprise applications will feature task-specific AI agents by 2026, up from less than 5\% in 2025.
\begin{ReviewerBEnv}International Data Corporation (IDC)\end{ReviewerBEnv}~\cite{IDC} also forecasts agent-based AI as a major driver of enterprise IT investment through large-scale agent orchestration.

%%%% ------------------
% OLD
%Over the past two years, we saw an increased practitioner demand for concrete guidance and shared experience. Newly established conferences dedicated to agentic AI have reported high participation. In parallel, established large-scale industry conferences have dramatically expanded their coverage of agentic systems; for example, recent editions of  \begin{ReviewerBEnv}Amazon Web Services (AWS)\end{ReviewerBEnv} re:Invent featured 720 sessions on AI agents within a single week, spanning architecture, tooling, deployment, and governance~\cite{reinvent25}.

% ROB Rephrased
Over the past two years, practitioner demand for concrete guidance and shared experience has increased significantly. Newly established conferences on agentic AI report strong participation, while major industry events have expanded their focus on agentic systems. For instance, recent editions of \begin{ReviewerBEnv}Amazon Web Services (AWS)\end{ReviewerBEnv} re:Invent featured 720 sessions on AI agents within a single week, covering architecture, tooling, deployment, and governance~\cite{reinvent25}.

%%%% ------------------

% While this explosion of practitioner-oriented material signals intense interest, it also creates a growing problem of information overload. Practitioners face a rapidly expanding volume of talks, frameworks, demos, and architectural advice fragmented across vendors, events, and communities. 

%Keeping up with this pace has become increasingly difficult, creating a growing problem of information overload,  particularly for teams operating under delivery pressure. As a result, architectural decisions about agent adoption and design are often made with limited consolidated evidence about what works in practice.

%To address this challenge, 
%%%% ------------------

% OLD
%This article adopts a practitioner-centered analysis to distill what actually works, what breaks, and how practitioners structure agent systems today.
%We analyze a large collection of 234 
%publicly available practitioner conference talks and technical presentations on AI agents using a structured qualitative analysis supported by \begin{ReviewerAEnv}LLMs\end{ReviewerAEnv} with humans in the loop. Rather than proposing new frameworks or formal theories, our goal is to {\em consolidate recurring themes}, {\em architectural strategies}, and {\em practical lessons} reported by practitioners actively building and deploying agentic systems.

%ROB Shrinked
This article adopts a practitioner-centered analysis to identify what works, what fails, and how agent systems are structured in practice. We analyze 234 publicly available practitioner conference talks and technical presentations on AI agents using a structured qualitative approach supported by \begin{ReviewerAEnv}LLMs\end{ReviewerAEnv} with humans in the loop. Rather than proposing new frameworks or formal theories, our goal is to {\em consolidate recurring themes}, {\em architectural strategies}, and {\em practical lessons} reported by practitioners building and deploying agentic systems.

Specifically, the article is guided by three objectives. First, we examine how organizations transition from early experimentation with AI agents to production adoption, focusing on \textit{motivations}, \textit{architectural factors}, and \textit{practitioner experience} influencing whether teams adapt existing systems or build agentic solutions from scratch. Second, we capture reusable architectural knowledge from industrial practice, identifying common \textit{strategies} and \textit{patterns} for designing agentic systems. Third, we investigate how agent-based architectures behave in real-world settings, highlighting \textit{differences across application domains} and the \textit{technologies} used to implement and operate \textit{LLM}-driven agentic systems.

%%%% ------------------

%OLD
% \begin{ReviewerAEnv}
% This article makes three main contributions. First, we provide an empirical analysis from the practitioner's perspective of AI agent adoption based on industry conference talks.
% Secondly, we synthesize the recurring architectural strategies and patterns used to build agentic systems in practice.
% Lastly, we analyze the agentic architecture implementation in specific domains and summarize the technologies and frameworks practitioners rely on to build or operate LLM-driven agentic systems.
% Together, these insights distill recurring lessons from practitioner experiences and provide practical guidance for engineers designing and deploying agentic systems.
% \end{ReviewerAEnv}

% ROB Rephrased
\begin{ReviewerAEnv}
This article makes three main contributions:
\begin{itemize}
    \item a large-scale empirical analysis of AI agent adoption grounded in 234 practitioner conference talks, providing insight into how organizations transition from experimentation to production;
    \item a consolidated view of recurring architectural strategies and patterns that characterize agentic systems in industrial practice;
    \item a cross-domain synthesis of agentic system implementations, identifying the technologies and frameworks used to build and operate \textit{LLM}-driven agents.
\end{itemize} 
Together, these contributions translate dispersed practitioner experience into structured knowledge and practical guidance for engineers designing and deploying agent-based systems.
\end{ReviewerAEnv}

% \todo[inline]{Write somewhere about the use of 'GXX' for references to groups of talks presents in the associated repository.\\Matteo: done in the end of the Resaerch Objectives section.}

% he introduction should provide background information... 
% \textcolor{blue}{I suggest here to write some background information about:
% \begin{itemize}
% \item {\bf The hype of AI Agents for industry:} something like... AI Agents are becoming a transformative force in software engineering. No longer limited to passive support tools, they increasingly act as autonomous, reasoning, and cooperative participants within development workflows. Whether generating source code, writing documentation, producing tests, or autonomously diagnosing and deploying systems, AI-driven agents are reshaping what software teams do and what they can achieve. As these agents begin to function as integral actors across the software lifecycle, software engineering enters a new era—one that poses significant opportunities and equally substantial challenges for both academia and industry.
% \item {\bf Some numbers:} in terms of conferences, talks, papers, etc
% \item {\bf the goal of this article:} analyze the trends from the practitioners lens, and by running practitioners video analysis.
% \item {\bf RQs:} we could add here the questions we want to answer through the video analysis. I would not call them research questions, though!
% \end{itemize}
% }

%\break 
%\vadjust{\pagebreak} 

%==================================
\section{AI AGENTS ARCHITECTURES}
%\todo[inline]{Henry and Davide}
%\textcolor{blue}{Since our focus is mostly, even if not completely, on AI agents' architecture, we could write a section here summarizing what an architecture for AI agents looks like, its main components, and ongoing research. If I have time, I might draft this section}.

%Recent discussions on the engineering of AI-based agents \cite{} reveal the mainstream role of architectures for generating quality systems.

%%% -----------------
%% OLD
%Agentic systems are opening new perspectives on how to design AI agents. This spans from architecting a single AI agent to architecting multi-agent AI applications. 

%%% Rob Shrinked
Agentic systems open new perspectives for designing AI agents, from single-agent to multi-agent applications.
%%% -----------------
In the former case,
new architecture design decisions are required to: use short-term or long-term memory strategies for storing, organizing, and retrieving context, experiences, or task-relevant knowledge; context compression or summarization for long‑running tasks to handle limited context windows; choosing which tools, \begin{ReviewerBEnv}Application Programming Interfaces (APIs)\end{ReviewerBEnv}, plugins, or actuators the agent can use to interact with the world beyond text; optimize for latency and cost by limiting unnecessary tool calls. Those decisions impact the internal structure of an AI agent, including its memory mechanisms, tools, planning and reasoning capabilities, and self-reflection. 

%%% ------------
%OLD
%When dealing with the architecture of multi-agent AI systems, decisions on how to intelligently divide labour based on skill and helpful feedback from a variety of agent personas, how to connect agents (e.g., through a choreography or an orchestration), and which communication patterns may be used in which case, may become fundamental to engineer quality AI-based systems.

%ROB Shrinked
When designing multi-agent AI systems, decisions on how to divide labor by skill and feedback across agent personas, how to connect agents (e.g., choreography or orchestration), and which communication patterns to use become fundamental for engineering quality AI systems.
%%% ------------

%(which may include personas, goals, memory, and an agent workflow together with the selected AI engine, and cooperate with external tools and other components) to architecting a multi-agent system (

%Properly building agentic systems requires new architectural ....

%
% architectural and conceptual foundations of AI Agents in SE

% Building reliable and trustworthy AI agents requires new architectural design, 

%==================================
\section{SIDEBAR: The Analysis Process}
\input{Sections/sidebar}
% \textcolor{blue}{I would suggest briefly reporting our protocol on a sidebar. This IEEE Sw article provides an example: https://tinyurl.com/32vn4b7n}

% \textcolor{blue}{Here is another example: https://tinyurl.com/y6d4sf9a}

%\begin{figure}
%\centerline{\includegraphics[width=18.5pc]{fig1.jpg}}
%\caption{Note that ``Figure'' is spelled out. There is a period after the figure number, followed by one space. It is good practice to briefly explain the significance of the figure in the caption. (From [``Title''],$^1$ used with permission.)}\vspace*{-5pt}
%\end{figure}

%==================================
\section{Research Objectives}
\input{Sections/findings}

%==================================
\section{TAKEAWAYS}
\input{Sections/takeaways}

\section{LIMITATION}
\begin{ReviewerAEnv}
%While the dataset of our analysis includes systematically filtered practitioner conference talks and technical presentations, it may not capture all architectural practices across the software industry. These talks may focus on successful cases or vendor-driven narratives, and may not fully reflect industry experiences, such as small or failed cases.
%In addition, because our analysis relies on LLM-enabled approaches, LLMs may generate hallucinations or inaccurate information, influencing the final results~\cite{mohammadabadi2025survey}. Although we have mitigated this risk through multi-model validation and manual review~\cite{su2025emerging}, it still exists.
While our dataset includes systematically filtered practitioner conference talks and technical presentations, it may not capture all architectural practices across the software industry. These sources may emphasize successful or vendor-driven cases and underrepresent smaller or failed efforts. In addition, as our analysis relies on LLM-enabled approaches, it may be affected by hallucinations or inaccuracies~\cite{mohammadabadi2025survey}. Although mitigated through multi-model validation and manual review~\cite{su2025emerging}, this risk remains.
\end{ReviewerAEnv}

%==================================
\section{CONCLUSION}
\input{Sections/conclusions}

\section{ONLINE MATERIALS}
\label{sec:onlineApdx}
All supplementary materials are available in the online appendix, including transcripts, LLM scripts, inclusion and exclusion criteria, the complete codebook, and \ReviewerB{extended methodology and linguistic and textual analysis of transcripts}~\cite{onlineappendix2026}.

% All supplementary materials are available in the online appendix, including the talk transcripts, the LLM scripts, the inclusion/exclusion criteria, the complete codebook, \ReviewerB{extended methodology, and transcripts linguistic and textual properties analysis}~\cite{onlineappendix2026}.
% \footnote{\url{https://doi.org/10.5281/zenodo.18154316}}.

%==================================
\section{ACKNOWLEDGMENTS}
This work has been funded by the Research Council of Finland (grants n. 359861 and 349488 - MuFAno) and Business Finland (grant 6GSoft~\cite{akbar_6gsoft_2024}), by FAST, the Finnish Software Engineering Doctoral Research Network, funded by the Ministry of Education and Culture, Finland, and with hardware support by CSC, the Finnish IT Center for Science.

% \section{Declaration of generative AI and AI-assisted technologies in the writing process}
% During the preparation of this work, the authors used ChatGPT to improve language and readability. After using this service, the authors reviewed and edited the content as needed and take full responsibility for the content of the publication.

%==================================

%==================================

\input{Sections/bios}

\def\refname{REFERENCES}

\bibliographystyle{IEEEtran}
\bibliography{references}

\end{document}

%% file: preamble.tex
\usepackage{graphicx}
\usepackage[colorlinks,urlcolor=blue,linkcolor=blue,citecolor=blue]{hyperref}
\expandafter\def\expandafter\UrlBreaks\expandafter{\UrlBreaks\do\/\do\*\do\-\do\~\do\'\do\"\do\-}
\usepackage{upmath,color}

\usepackage{todonotes}
\usepackage{tikz}
\usetikzlibrary{arrows.meta,positioning}
\usepackage{graphicx}
\input{RevisionsEnvs}

\jvol{XX}
\jnum{XX}
\paper{8}
\jmonth{Month}
\jname{Publication Name}
\jtitle{Publication Title}
\pubyear{2021}

\usepackage[dvipsnames]{xcolor}
\usepackage{textcomp}

%% file: RevisionsEnvs.tex
\newif\ifshowReviewColors
\showReviewColorsfalse   % set to \showReviewColorsfalse to disable colors

\usepackage[table]{xcolor} % "table" option recommended
\usepackage{colortbl}      % for \arrayrulecolor

\definecolor{ReviewerAColor}{RGB}{0,128,128} % teal
\definecolor{ReviewerBColor}{RGB}{200,0,200} % magenta
\definecolor{ReviewerCColor}{RGB}{220,0,0}   % red

\newcommand{\ReviewerB}[1]{%
  \ifshowReviewColors
    \textcolor{red}{#1}%
  \else
    \textcolor{black}{#1}%
  \fi
}

\newenvironment{ReviewerAEnv}
  {%
    \begingroup
    \ifshowReviewColors
      \color{magenta}%
      \arrayrulecolor{magenta}% table borders in same color
    \else
      \color{black}%
      \arrayrulecolor{black}%
    \fi
  }
  {%
    \arrayrulecolor{black}% restore default for following tables
    \endgroup
  }

\newenvironment{ReviewerBEnv}
  {%
    \begingroup
    \ifshowReviewColors
      \color{red}%
      \arrayrulecolor{red}%
    \else
      \color{black}%
      \arrayrulecolor{black}%
    \fi
  }
  {%
    \arrayrulecolor{black}%
    \endgroup
  }

%% file: Sections/sidebar.tex
\begin{mdframed}[hidealllines=true,backgroundcolor=Dandelion]

% \todo[inline]{Rob: do we have to specify here the conferences/talks selection criteria used? Or at least mention the conferences considered?}

Our study followed a lightweight, practitioner-oriented qualitative analysis process designed to extract architectural insights from a large collection of recorded industry talks and video transcripts. The goal was not to build exhaustive theory, but to capture recurring patterns, strategies, and challenges communicated by practitioners when adopting and operating AI agent-based architectures. We applied a recent methodology for applying LLM in qualitative analysis~\cite{robredo2025were, su2025emerging}.

% \subsection{Data Collection}
% We collected publicly available practitioner talks, conference presentations, and long-form technical briefings published between 2023--2025. All videos were transcribed automatically and then manually cleaned to ensure fidelity of technical terminology. The dataset represents a diverse mix of engineering domains, including cloud platforms, autonomous systems, and enterprise AI tooling.

\subsection{Data Collection}

% \todo[inline]{Henry: this repeats what already presented in the other Data Collection paragraph! Matteo: it was inadvertelty a copy/paste issue}.

We gathered 234 public practitioner talks, technical briefings, and conference presentations released between September 2024–September 2025. \textcolor{black}{We selected the most relevant practitioner conference using four inclusion criteria: (i) at least five prior editions ($\geq 5$), (ii) talks delivered by practitioners or industry-affiliated researchers, (iii) sessions conducted in English, and (iv) full video recordings freely available online~\cite{su2025emerging}.} Videos were automatically transcribed and manually checked. An automated pipeline of Bash scripts was used for video transcription, comprising the following steps: i) video download using the open source \textit{yt-dlp} tool\footnote{https://github.com/yt-dlp/yt-dlp} ii) audio extraction and cut using \textit{ffmpeg}\footnote{https://www.ffmpeg.org/}, and iii) transcript extraction using \textit{Whisper-XXL}\footnote{https://github.com/Purfview/whisper-standalone-win/releases/tag/Faster-Whisper-XXL}, a Transformer-based deep learning \begin{ReviewerBEnv}automatic speech recognition (ASR)\end{ReviewerBEnv} model. %Whisper was configured to produce English-language transcriptions with plain text (.txt) as the output format. 

%%% ---------------
%OLD
%Thereby, four authors, working in pairs and with a final judge intervening in cases of disagreement, independently reviewed the collected video transcripts and, based on the inclusion and exclusion criteria, filtered the initial 234 transcripts reducing it to 138 transcripts.

% ROB Shrink
Four authors, working in pairs with a final judge resolving disagreements, independently reviewed the transcripts, applying inclusion and exclusion criteria to reduce 234 transcripts to 138.
%%% ---------------

%%%% rob revomed
%The material spans multiple sectors, including cloud services, automation platforms, and domain-specific AI deployments.

% \todo[inline]{Matteo: Add IC/EC table} 

%%%%% --------------
% OLD
%We used one \textbf{Large Reasoning Model (LRM)} as the primary engine for analyzing transcripts. We adopted a \begin{ReviewerBEnv}retrieval-argumented generation (RAG)\end{ReviewerBEnv} approach to handle the fact that some transcripts were too large for the LLM's context length. 

% ROB Shrink
We used a \textbf{Large Reasoning Model (LRM)} as the primary engine for transcript analysis, adopting a \begin{ReviewerBEnv}retrieval-argumented generation (RAG)\end{ReviewerBEnv} approach to handle transcripts exceeding the LLM’s context length.
%%%%% --------------
\begin{ReviewerBEnv}The LRM processed each transcript and produced structured outputs that captured architectural statements such as adoption drivers, coordination patterns, workflow structures, and enabling technologies.\end{ReviewerBEnv} The LRM’s role mirrors its function in prior architecture compliance studies, where it acts as the main reasoning component for evidence extraction and classification (thematic and axial coding). 

% \todo[inline]{Henry: this text does not highlight the enormous human in the loop effort!} 

To ensure accuracy and reduce single-model bias, we employed \textbf{three independent Validator models (V1, V2, V3)}. Each Validator reviewed the LRM’s output and checked it against the original transcript segments:
\begin{itemize}
    \item whether the LRM’s summaries faithfully reflected the transcript,
    \item whether architectural concepts were overstated or hallucinated,
    \item and whether the extracted insights maintained practitioner intent.
\end{itemize}

\begin{ReviewerBEnv}Validators produced a simple agree/disagree decision with comments. Only insights confirmed by at least two Validators were retained. \end{ReviewerBEnv}
% This mirrors the multi-judge structure used in the ADR study, where multiple independent LLMs stabilize classification decisions and prevent model drift.

Finally, according to Su et al.~\cite{su2025emerging} and Robredo et al.~\cite{robredo2025were}, \begin{ReviewerBEnv}two authors reviewed the consolidated LRM+Validator outputs and assessed the LLM identified themes. Additional details, including prompts, scripts, extended methodology, and full coding schema, are provided in the online appendix~\cite{onlineappendix2026}.\end{ReviewerBEnv}

% \subsection{Outcome of the Method}
% This process resulted in a compact set of architectural adoption insights, design strategies, structural patterns, and technology categories directly grounded in practitioner experience. The approach prioritizes interpretability and practical relevance, making the findings suitable for architects evaluating or planning AI agentic systems in industrial settings.

% \todo[inline]{June: Should we have a Figure of the whole methodology in the associated repo to refer to?}

\end{mdframed}

%% file: Sections/findings.tex
%%%----------
%%%OLD
%Organizations exploring AI agent-based architectures face three fundamental challenges: understanding how such systems are adopted, determining how they should be architected, and assessing their performance once deployed. Our study addresses these challenges through three structured objectives. 

%%% ROB Shrink
Organizations exploring AI agent-based architectures face three challenges: understanding adoption, designing architectures, and assessing performance after deployment. Our study addresses them through three objectives.
%%%----------

\textbf{Objective~1} examines how companies adopt agent-based architectures, focusing on motivations, the architectural factors, and the practitioner experience that influence whether teams migrate existing systems or build agentic solutions from scratch. 

\textbf{Objective~2} captures the reusable architectural knowledge emerging from industrial practice, clarifying how practitioners define agents, assign responsibilities, and organize coordination within and across multi-agent setups. 

 \textbf{Objective~3} investigates how agent-based architectures are implemented and applied in real-world settings, highlighting differences across application domains and the technologies used to build and operate LLM-driven agentic systems.

 In the following sections, we use \textbf{G$x$} to denote a group of talks that provide evidence for a specific claim. Due to space constraints, the full mapping from group IDs to talks is available in the replication package.

 \begin{ReviewerBEnv}
     All in all, the transcripts average ~4k words and ~23k characters, with moderate lexical diversity (TTR = 0.33, MTLD = 64.73) and relatively stable syntactic complexity (MDD = 2.62), while showing substantial variability and right skewness.
 \end{ReviewerBEnv}
 
 % Together, these objectives offer a coherent, practitioner-focused view of how agent architectures evolve from initial pilots into scalable, reliable production systems.

\section{Understanding How Companies Adopt Agent-Based Architectures}

% \todo[inline]{Matteo: - Add Line IC/EC, @Davide add lines to specify cofnernece seleciotion, @Matteo fix Timeline figure and add lines. }

\input{Figures/timeline}

%%%% -----------
% OLD
% Many teams are experimenting with AI agents, but the roadmap from early pilots to meaningful architectural change is far from clear. \textit{Objective~1} looks at how organizations actually move through this journey. We examine the \textit{motivations} behind adopting agentic patterns, 
% the \textit{architectural factors} that influence whether practitioners choose to develop an agent-based solution from scratch or migrate an existing system, 
% and the \textit{practitioner experience} summarizing the practical challenges of adoption, from toolchain limitations and engineering complexity to reliability and trust in production. We also analyze agent adoption over time; Figure~\ref{fig:agent_timeline} briefly shows the timeline due to space constraints.

% ROB Shrink
Many teams are experimenting with AI agents, but the path from early pilots to meaningful architectural change remains unclear. \textit{Objective~1} examines how organizations progress along this journey. We analyze the \textit{motivations} for adopting agentic patterns, the \textit{architectural factors} influencing whether to build from scratch or migrate existing systems, and the \textit{practitioner experience}, including challenges such as toolchain limitations, engineering complexity, and reliability and trust in production. We also analyze adoption over time; Figure~\ref{fig:agent_timeline} summarizes the timeline.
%%%% -----------

\subsection{Motivation} 
Practitioners consistently point to a mix of technological and organizational forces driving the adoption of agent-based architectural patterns. Teams are motivated by the {\em automation potential} enabled by improved reasoning, tool use, and extended autonomous operation, which together reduce manual effort and increase productivity (\textbf{G1}). {\em Cost reductions} in frontier model inference make agent-based automation economically competitive with manual processes (\textbf{G2}). In parallel, {\em engineering constraints} such as scalability, latency, reliability, and enterprise interoperability make agent architectures an attractive fit for modern industrial systems (\textbf{G3}). Practitioners also emphasize {\em customer-facing} benefits, including more personalized, predictive, and proactive user experiences (\textbf{G4}). Strategic {\em organizational pressure} further supports adoption, as senior leadership often encourages teams to restructure products and architectures around intelligent agent capabilities (\textbf{G5}).

\subsection{Architectural factors} 
Talks highlighted seven key architectural considerations that influence whether AI agent systems are built from scratch or migrated from existing non-agent architectures. First, {\em planning, reasoning, and grounding} requirements matter. When current systems cannot support agent planning, complex reasoning, search-based action selection, or robust real-world grounding, migration demands heavy refactoring. It often pushes teams toward greenfield development (\textbf{G6}). Second, ease of {\em integration} and {\em extensibility} can favor migration. Architectures that support incremental evolution allow teams to plug in custom agents via APIs while reusing familiar components, tools, and infrastructure (\textbf{G7}). Third, standardized communication and the maturity of the \begin{ReviewerBEnv}model context protocol (MCP)\end{ReviewerBEnv} influence feasibility. Mature {\em agent communication protocols} and MCP-based tooling make migration more attractive, whereas limited or immature protocol infrastructure can push practitioners to design entirely new agent architectures (\textbf{G8}). 

Fourth, {\em simplicity} in build, test, and release engineering lowers migration friction. Clear abstraction layers, separation of concerns, and supervised frameworks help teams evolve existing systems while keeping build, test, and release processes stable (\textbf{G9}). Fifth, the balance between {\em workflow control and autonomy} is critical. static, predefined workflows scale well for predictable tasks, but highly autonomous behavior for unpredictable work often requires substantial re-architecting, making a fresh design more appealing (\textbf{G10}). Sixth, the interaction pattern fit plays a role. Migration is easier when current systems already support the required interaction patterns, such as natural language interfaces and programmatic control; otherwise, bigger architectural changes are needed, which can favor starting from scratch (\textbf{G11}). Finally, {\em unified entity abstraction} is a decisive factor. Architectures that treat humans, tools, and language models as a shared type of entity simplify coordination and reasoning. Systems that lack this unified view often require extensive refactoring, making rebuilding the more pragmatic option (\textbf{G12}).

\subsection{Practitioner Experience}
Several challenges shape how organizations adopt agents in practice. Many teams expected rapid acceleration once agents demonstrated initial value, yet early deployments exposed instability and gaps in toolchains, which highlighted the need for deeper evaluation, monitoring, and framework comparison (\textbf{G13}). End-to-end engineering complexity also posed obstacles. Teams iterated repeatedly on architecture designs, encountered {\em integration issues} across tools, frameworks, and memory, and faced infrastructure scalability pressures that required coordinated hardware-software evolution and ``glue'' toolkits for observability and workflow management (\textbf{G14}). 

{\em Use-case selection} emerged as another critical factor. Several agent concepts failed due to weak alignment with business needs, which pushed practitioners to adopt clearer selection criteria and to combine traditional \begin{ReviewerBEnv}Machine Learning (ML) and Generative AI (GenAI)\end{ReviewerBEnv} when appropriate (\textbf{G15}). {\em Reliability and trust} further influenced adoption strategies. Latency, hallucinations, and error rates affected user confidence, so teams increasingly relied on staged rollout playbooks that start with simple tasks, gradually expand autonomy, enforce comprehensive testing, and incorporate continuous production feedback (\textbf{G16}).

\section{Architectural Strategies and Patterns}

Teams building agent-based solutions often ask the same practical questions: How many agents do we need? What responsibilities should each agent take? How should agents coordinate? %\textit{Objective~2} focuses on answering these questions by identifying reusable \textit{architectural strategies} that practitioners can depend on. %We look at how agents are \textit{decomposed}, how \textit{roles} are assigned, and which \textit{communication and coordination patterns} recur across industrial systems. We summarize the identified strategies and patterns in Figure~\ref{fig:objective2_two_column}.
\textit{Objective~2} addresses these questions by identifying reusable \textit{architectural strategies}. We examine how agents are \textit{decomposed}, how \textit{roles} are assigned, and which \textit{communication and coordination patterns} recur across industrial systems. These strategies and patterns are summarized in Figure~\ref{fig:objective2_two_column}.

\input{Figures/O2}

\subsection{Strategies} Talks highlighted a diverse set of \textbf{architectural strategies} that guide practitioners in structuring and organizing multi-agent AI systems. A common starting point is \textbf{task decomposition} and \textbf{role specialization}, where complex activities are broken into manageable subtasks and assigned to agents with clearly defined responsibilities. Thus, reducing the system complexity improves execution precision (\textbf{G17}). Many systems rely on \textbf{supervisor-orchestrated} and \textbf{graph-structured coordination}, in which coordinator agents interpret goals, allocate and validate tasks, and manage dependencies through graph-based interaction frameworks (\textbf{G18}). Practitioners also adopt \textbf{incremental growth} and \textbf{agent count budgeting}, beginning with a minimal set of agents and adding sub-intelligences only when performance gains are measurable, helping prevent coordination failures and illusion compounding (\textbf{G19}). Collaboration is frequently enabled through \textbf{shared context} and \textbf{unified communication protocols}, which ensure agent interoperability and support scalable teamwork (\textbf{G20}). Some architectures introduce \textbf{dynamic team composition} and \textbf{role reassignment}, where sub-agents can be created or reorganized at runtime to meet workflow needs, quality constraints, or changing environments (\textbf{G21}).

Other strategies align agent roles with organizational structures through \textbf{personal/department-aligned role mapping}, matching agents to roles such as sales, HR, or IT to maintain real operational fidelity (\textbf{G22}). Reliability-driven approaches include \textbf{expert ensembles} and \textbf{majority decision aggregation}, where multiple evaluators operate in parallel and their judgments are combined to improve robustness (\textbf{G23}). Several systems employ \textbf{feedback loops} and \textbf{reinforcement-based self-improvement} to allow agents to refine their behavior through iterative correction cycles (\textbf{G24}). In addition, \textbf{tool-centric role design} focuses on building precise toolkits—such as analysis, repair, or recovery—rather than increasing agent intelligence, improving controllability, and reducing token consumption (\textbf{G25}). Some systems emphasize a \textbf{minimal core structure} for agents while relying on \textbf{structured environment support}, enabling autonomy without premature rigidity (\textbf{G26}). Finally, \textbf{multi-level local–global agent layers} divide responsibilities across tiers, using registries and access controls to govern cross-cluster and cross-scope interactions (\textbf{G27}).

\subsection{Patterns}
Talks identified 10 \textbf{architectural patterns} that shape both single-agent design and multi-agent coordination, organized into four recurring categories: coordination, control, reliability, and evolution. 

\textit{Coordination patterns} dominate practitioner discussions and address how agents interact and distribute work. A foundational pattern is the \textbf{iterative reasoning–action–observation loop}, in which agents follow cyclical decision-making processes or structured exploration strategies, such as reconnaissance, enumeration, or knowledge graph construction (\textbf{G28}). Coordination is further supported through \textbf{workflow-oriented static or dynamic collaboration}, enabling agents to execute predefined sequences or to operate autonomously beyond LLM-driven action selection (\textbf{G29}). At scale, practitioners rely on \textbf{message-based asynchronous communication} and \textbf{microservice deployment}, using queues and event-driven interactions to support retries, long-running tasks, and distributed execution (\textbf{G32}). More structured coordination emerges through \textbf{centralized skill and function registries} and \textbf{inter-agent protocol coordination}, which enable synchronous or asynchronous capability discovery and invocation via unified routing and versioned toolchains (\textbf{G33}). \textbf{Graph-based coordination} further models tasks as dependent nodes, enabling precise routing, dependency resolution, and domain-specific dispatching (\textbf{G34}). Across systems, multiple coordination paradigms coexist, including \textbf{hierarchical supervision}, \textbf{coordinator–executor models}, \textbf{teamwork}, and \textbf{swarm or hybrid patterns}, each balancing autonomy and oversight in different ways (\textbf{G35}).

\textit{Control patterns} focus on governing agent behavior and the scope of interactions. \textbf{Nested dialogue structures} and \textbf{embedded agent interactions}, such as serialized group chats or agents invoking other agents, allow practitioners to constrain reasoning flows while supporting multi-step coordination (\textbf{G36}). At enterprise scale, a \textbf{centralized coordination control plane} enforces global policies, manages shared state, supports auditing, and dynamically routes resources, forming the backbone of large multi-agent deployments (\textbf{G37}).

\textit{Reliability patterns} address stability, safety, and compliance. \textbf{Multimodal memory architectures} integrate working, long-term, and collective memory to maintain continuity of reasoning and enable robust cross-agent information sharing (\textbf{G30}). In parallel, \textbf{policy-driven safeguards} and \textbf{remediation mechanisms} are introduced to prevent behavioral drift, detect failures, and ensure compliance across agent workflows (\textbf{G31}).

Finally, \textit{evolution patterns} support long-term adaptability. By combining modular coordination mechanisms, extensible registries, and layered control infrastructures, these patterns allow agent systems to evolve incrementally as tasks, environments, and organizational requirements change, without requiring disruptive architectural rewrites.

\section{Real-World Implementation and Architectural Implications of Agentic Systems}

%%% -------------
% OLD
% Practitioners want to know what works, for whom, and under which conditions. \textit{Objective~3} investigates the reported \textit{impacts} of adopting agent-based architectures, looking at \textit{qualities} that matter in industry, such as scalability, adaptability, and maintainability. We also compare \textit{architectural practices} across domains and identify the technologies companies rely on to build LLM-enabled agentic systems.

% ROB Shrink
Practitioners want to know what works, for whom, and under which conditions. \textit{Objective~3} examines the \textit{impacts} of agent-based architectures, focusing on key \textit{qualities} such as scalability, adaptability, and maintainability. We compare \textit{architectural practices} across domains and identify the technologies used to build LLM-enabled agentic systems.
%%% -------------

\subsection{Application Domain Differences}
Talks explored several perspectives on how communicated architectural practices for AI agents differ across domains and tasks. A first difference concerns \textbf{interoperability setup through discovery and standardized protocols}. Domains vary in how agents discover each other and publish capabilities, ranging from machine-readable capability documentation and contextual discovery to stricter use of inter-agent protocols that support secure and explainable coordination (\textbf{G38}). A second difference lies in \textbf{task-execution ``interaction contracts''}, in which domains shape how agents negotiate task-critical parameters, shared data formats, expected outputs, and the choice between flexible natural-language interaction and precise programming-language interfaces (\textbf{G39}). A third difference involves \textbf{control-versus-autonomy choices} in user experience and workflow design. End-to-end automation often employs tightly controlled, predefined workflows. At the same time, interactive assistants emphasize dynamic and more autonomous flows that remain appropriate for well-bounded tasks rather than highly uncertain, open-ended scenarios (\textbf{G40}). A fourth difference concerns \textbf{enterprise process encoding as \begin{ReviewerBEnv}directed acyclic graph (DAG)\end{ReviewerBEnv} workflows}. Many business settings regulate agent execution using structured, DAG processes in which predefined prompts encode workflow steps, tool integrations, and execution order in accordance with domain-specific constraints (\textbf{G41}). 

Further differences emerge between digital and physical environments and among specialized domains. In digital domains, \textbf{observe–plan–act implementations} often rely on \begin{ReviewerBEnv}hypertext markup language (HTML)\end{ReviewerBEnv} or visual page parsing, manipulation, and API calls, whereas physical robots depend on perception pipelines, semantic scene understanding, and motion control, which leads to substantially different architectural realizations across stages of the same meta-architecture (\textbf{G42}). In cybersecurity-focused systems, \textbf{cybersecurity-specific practices} evolve through iterative trial-and-error. Agents frequently alternate between iterative command-adjustment cycles and structured scenario construction, for example, through reconnaissance and knowledge-graph building, under strong efficiency, cost, and operational constraints (\textbf{G43}).

\subsection{Technologies}
Talks also described ten categories of \textbf{technologies} used to implement LLM-based agentic systems. A first category is the \textbf{LLM portfolio and serving runtimes}. Industrial systems typically integrate multiple frontier and domain-specific models, such as GPT-4, GPT-4O, Claude, Gemini, Qwen, DeepSeek, Grok-3 mini, and Forward mini, exposed via embedded endpoints, inference endpoints, and reorderers with security protections. Some deployments also rely on local or open-source runtimes, including OLAMA, LMStudio, \begin{ReviewerBEnv}Vectorized Large Language Model Serving System (vLLM), Structured Generation Language (SGLang)\end{ReviewerBEnv}, and container-based inference stacks (\textbf{G44}). A second category comprises \textbf{agent frameworks and orchestration platforms} that support multi-agent workflows, planning, tool use, memory, and code generation, using frameworks such as AutoGen, Magentic1, Line Graph and its derivatives, Crew AI, LangGraph, Goose, Land Graph, \begin{ReviewerBEnv}Agent Development Kit (ADK)\end{ReviewerBEnv}, Autogens, Pedantic AI, Strands \begin{ReviewerBEnv}Software Development Kit (SDK)\end{ReviewerBEnv}, and Amazon Bedrock Agent Core (\textbf{G45}). A third category is \textbf{tool use and protocol-based integration}. These stacks center on APIs and custom functions, standardized integration standards such as MCP, tool invocation protocols, connection adapters, and related components, including tool servers, API agents, diagnostic tools, and MCP tool catalogs (\textbf{G46}). A fourth category includes \textbf{memory, context, vector search, and RAG services}, where short- and long-term memory, session history, context stores, knowledge bases, vector databases, and semantic search are combined into RAG-as-a-Service style components (\textbf{G47}). A fifth category focuses on \textbf{planning methods and workflow structuring}, using React-style loops, chains or trees of reasoning, DAG-based workflows, and pre-execution planning integrated with workflow engines for serialized, dependency-aware execution (\textbf{G48}).

Moreover, \textbf{runtime deployment and distributed execution} typically use Kubernetes and microservice architectures, serverless inference, and distributed execution models, sometimes extended with edge–cloud collaboration, with platforms like Amazon Bedrock Agent Core as deployment backbones (\textbf{G49}). \textbf{Hardware acceleration and specialized compute} stacks emphasize dedicated inference hardware, including \begin{ReviewerBEnv}Graphics Processing Units (GPUs)\end{ReviewerBEnv}.

%% file: Figures/timeline.tex
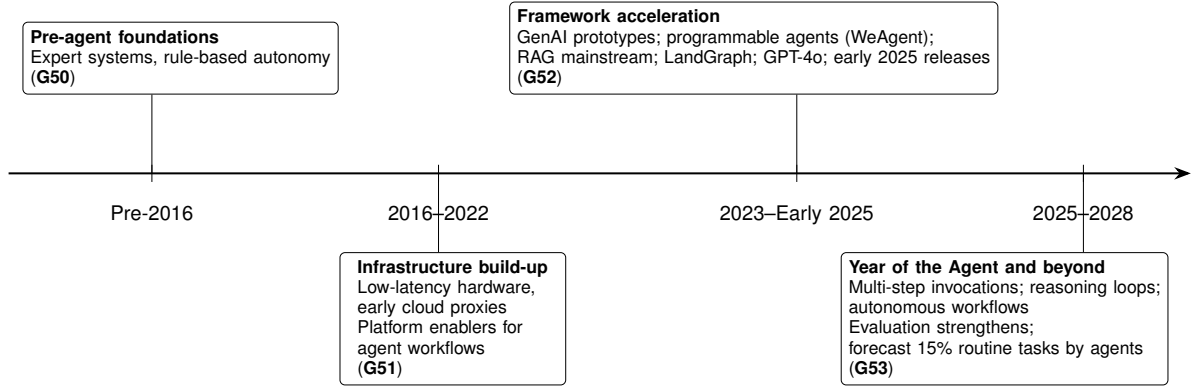
\begin{figure*}[tb]
\centering
\resizebox{\linewidth}{!}{
\begin{tikzpicture}[
    >=Stealth,
    timeline/.style={thick,-{Stealth}},
    box/.style={rectangle, rounded corners=2pt, draw, align=left, font=\scriptsize,
                minimum width=3.15cm, inner sep=3pt},
    year/.style={font=\footnotesize}
]

% --- Timeline bounds (more left breathing room) ---
\def\xStart{-2.0}
\def\xEnd{14.5}

% Horizontal timeline
\draw[timeline] (\xStart,0) -- (\xEnd,0);

% Year ticks
\foreach \x/\label in {
    0/Pre-2016,
    4/2016--2022,
    9/2023--Early 2025,
    13/2025--2028
} {
    \draw (\x,0.12) -- (\x,-0.12);
    \node[year, below=0.22cm] at (\x,-0.12) {\label};
}

% --- Boxes (slightly shifted LEFT to re-center) ---

% Pre-2016 (Up) --- was 0.6, now 0.4
\node[box, above=1.1cm of {(0.4,0)}] (b1) {
    \textbf{Pre-agent foundations}\\
    Expert systems, rule-based autonomy\\
    {(\textbf{G50})}%[126, 152]
};

% 2016–2022 (Down) --- was 4.4, now 4.2
\node[box, below=1.1cm of {(4.2,0)}] (b2) {
    \textbf{Infrastructure build-up}\\
    Low-latency hardware,\\ early cloud proxies\\
    Platform enablers for\\ agent workflows\\
    {(\textbf{G51})}%[13]
};

% 2023–Early 2025 (Up) --- was 8.6, now 8.4
\node[box, above=1.1cm of {(8.4,0)}] (b3) {
    \textbf{Framework acceleration}\\
    GenAI prototypes; programmable agents (WeAgent);\\
    RAG mainstream; LandGraph; GPT-4o; early 2025 releases\\
    {(\textbf{G52})}%[90, 152, 13, 16, 86, 151]}
};

% 2025–2028 (Down) --- was 12.1, now 11.9
\node[box, below=1.1cm of {(11.9,0)}] (b4) {
    \textbf{Year of the Agent and beyond}\\
    Multi-step invocations; reasoning loops;\\ autonomous workflows\\
    Evaluation strengthens;\\ forecast 15\% routine tasks by agents\\
    {(\textbf{G53})}%[24, 117, 86, 90, 30]}
};

% --- Perfectly vertical connectors (perpendicular to axis) ---
\draw[thin] (0,0) -- (0,0 |- b1.south);
\draw[thin] (4,0) -- (4,0 |- b2.north);
\draw[thin] (9,0) -- (9,0 |- b3.south);
\draw[thin] (13,0) -- (13,0 |- b4.north);

\end{tikzpicture}
}

\caption{Timeline of industrial adoption phases for AI agent-based architectures.}
\label{fig:agent_timeline}
\end{figure*}

%% file: Figures/O2.tex
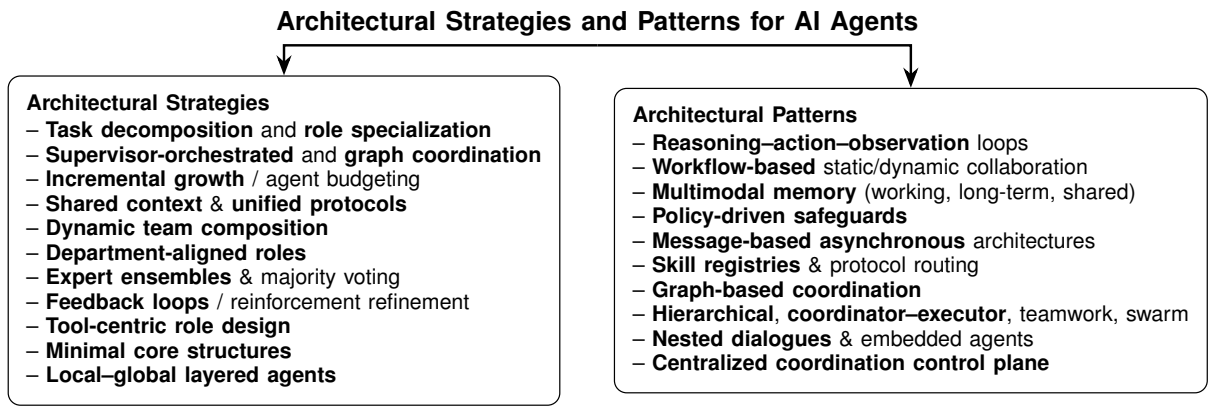
\begin{figure*}[t]
\centering
\resizebox{\textwidth}{!}{%
\begin{tikzpicture}[
    >=Stealth,
    title/.style={font=\small\bfseries},
    box/.style={rectangle, rounded corners=4pt, draw, align=left, font=\scriptsize,
                minimum width=6cm, inner sep=6pt},
    arrow/.style={->, thick}
]

% Title node at the top center
\node[title] (obj2) at (0,2.0) {Architectural Strategies and Patterns for AI Agents};

% Left box: Architectural Strategies
\node[box] (strategies) at (-3.6,-0.5) {%
    \textbf{Architectural Strategies}\\[1pt]
    -- \textbf{Task decomposition} and \textbf{role specialization}\\
    -- \textbf{Supervisor-orchestrated} and \textbf{graph coordination}\\
    -- \textbf{Incremental growth} / agent budgeting\\
    -- \textbf{Shared context} \& \textbf{unified protocols}\\
    -- \textbf{Dynamic team composition}\\
    -- \textbf{Department-aligned roles}\\
    -- \textbf{Expert ensembles} \& majority voting\\
    -- \textbf{Feedback loops} / reinforcement refinement\\
    -- \textbf{Tool-centric role design}\\
    -- \textbf{Minimal core structures}\\
    -- \textbf{Local--global layered agents}
};

% Right box: Architectural Patterns
\node[box] (patterns) at (3.6,-0.5) {%
    \textbf{Architectural Patterns}\\[1pt]
    -- \textbf{Reasoning--action--observation} loops\\
    -- \textbf{Workflow-based} static/dynamic collaboration\\
    -- \textbf{Multimodal memory} (working, long-term, shared)\\
    -- \textbf{Policy-driven safeguards}\\
    -- \textbf{Message-based asynchronous} architectures\\
    -- \textbf{Skill registries} \& protocol routing\\
    -- \textbf{Graph-based coordination}\\
    -- \textbf{Hierarchical}, \textbf{coordinator--executor}, teamwork, swarm\\
    -- \textbf{Nested dialogues} \& embedded agents\\
    -- \textbf{Centralized coordination control plane}
};

% Arrows from title to each box (L-shaped, clean)
\draw[arrow] (obj2.south) -| (strategies.north);
\draw[arrow] (obj2.south) -| (patterns.north);

\end{tikzpicture}%
}
\caption{Compact overview of key architectural strategies and structural patterns identified for Objective 2.}
\label{fig:objective2_two_column}
\end{figure*}

%% file: Sections/takeaways.tex
% \todo[color=green!40,inline]{Do we want to match takeways with talks groups and/or objectives?}

%\textcolor{blue}{We could run here a horizontal analysis on all the individual findings}
%\textcolor{purple}{June: Rewrite the formulation with the style of 'if you are using agentic ai, you should...'.}

The following takeaways distill recurring lessons from practitioner experiences into actionable guidance for engineers designing and deploying agentic systems.

%% initial version
%\textbf{agentic systems are usable today, but still require careful handling.}
%Recent progress in reasoning, memory, and orchestration has made building real agent based systems rather than just prototypes possible. However, early adopters report that reliability remains uneven and that evaluation tooling is still catching up. Treating reliability as a continuous engineering effort, supported by testing, monitoring, and cautious rollout strategies rather than assuming stability will emerge on its own, seeems to be the path to success.
%% version practitioner style
%\textbf{If you are using agent based systems today, you should treat them as usable but not yet self stabilizing technologies.} Advances in reasoning, memory, and orchestration now allow teams to build real systems rather than isolated demos. However, early adopters consistently report uneven reliability and immature evaluation tooling. Successful teams do not assume that stability will emerge automatically. Instead, they approach reliability as an ongoing engineering concern, supported by systematic testing, monitoring, and cautious rollout strategies.

% short takeaway version
\textbf{Treat agent based systems as usable but not self stabilizing technologies.}
If you are adopting agent based systems today, you should expect uneven reliability and immature evaluation tooling. Stability does not emerge automatically and requires continuous attention through testing, monitoring, and cautious rollout practices.

%% initial version
%\textbf{System engineering and integration as first challenges.}
%Practitioners consistently report that the most demanding aspects of building agentic systems arise from their software engineering foundations rather than from model behaviour. Integrating tools and APIs, designing memory and workflow layers, establishing orchestration mechanisms, and implementing observability and debugging pipelines consume a large fraction of development effort. This reinforces the view that agentic systems must be treated as engineered software artefacts, not isolated model wrappers.
%% version practitioner style
%\textbf{When building agentic systems, you should expect software engineering and integration to be the primary challenges, not model behavior.} Experiences shared by teams show that most effort goes into integrating tools and APIs, designing memory and workflow layers, setting up orchestration, and building observability and debugging pipelines. This reinforces a key lesson: agentic systems must be engineered as full software systems, not treated as thin wrappers around models.

% short takeaway version
\textbf{Plan for system engineering and integration as the main source of effort.}
When building agentic systems, you should allocate most of your resources to integrating tools and APIs, designing memory and workflow layers, implementing orchestration, and setting up observability and debugging. These engineering tasks dominate development time more than model behavior.

%% initial version
%\textbf{Structured architectural design to enhance effectiveness.}
%When the basic engineering work is in place, the performance and usefulness of an agentic system depend heavily on how its components are organised. Architectures that rely on clear task decomposition, modular roles, shared context, and well defined workflows tend to scale more effectively and adapt more easily to new requirements. These systems are also easier to maintain because responsibilities and information flows are explicit. In practice, many of the reported gains in scalability, adaptability, and maintainability appear only when the agentic architecture is designed with this kind of structure in mind.
%% version practitioner style
%\textbf{If you want your agentic system to scale and remain useful over time, you should invest in a structured architectural design.} In practice, teams achieve better results when systems rely on clear task decomposition, modular agent roles, shared context, and well defined workflows. These structures make systems easier to extend and maintain because responsibilities and information flows are explicit. Many of the reported gains in scalability and adaptability only emerge once this architectural structure is deliberately designed.

% short takeaway version
\textbf{Use explicit architectural structure to achieve scalability and maintainability.}
Agentic systems perform better when you design them around clear task decomposition, modular roles, shared context, and well defined workflows. Without this structure, many scalability and adaptability benefits fail to materialize.

\textbf{Adapt agentic architectures to the target domain.}
You should tailor autonomy, control, and interaction patterns to your domain constraints. Enterprise workflows, interactive systems, physical robots, and cybersecurity applications each require different architectural choices, and agentic designs do not generalize uniformly across these settings.

%\textbf{ Predictability and controlled autonomy underpin user trust.}
%User acceptance depends on latency, behavioural consistency, and clarity of system boundaries. Deployments that begin with constrained autonomy and familiar interaction patterns tend to retain user confidence, enabling autonomy to increase only after reliability is demonstrated. This staged approach aligns with established practices for introducing adaptive components into operational systems, where trust must be earned gradually.

\textbf{Build a Control Plane You Can Trust.} Standardize your agent integrations with mature protocols (e.g., MCP), then invest in a control plane that gives you routing, monitoring, evaluation, and policy enforcement. Add governance (roles, auditability, and compliance workflows) to deploy and scale agent systems safely, interoperably, and with confidence.

%% file: Sections/conclusions.tex
% \textcolor{blue}{conclusions}
% AI agents are moving from experimentation to everyday use. Practitioner accounts reveal recurring architectural strategies, coordination patterns, and integration challenges that influence how these systems behave in production. Together, these observations highlight the role of architectural structure and domain context in shaping agentic systems in practice.

% Further studies are needed to examine how these practices evolve as agent technologies mature and are adopted at larger scale.

% AI agents are moving from experimentation to everyday use. Practitioner accounts reveal how organizations progress from pilots to production adoption and what factors influence architectural decisions along that journey (\textit{Objective 1)}. Across industries, recurring coordination strategies and reusable patterns show how practitioners structure agentic systems for scalability and maintainability (\textit{Objective 2}). At the same time, reported differences across domains highlight how technologies and implementation choices shape real-world agent behavior and operational effectiveness (\textit{Objective 3}). Further studies are needed to examine how these practices evolve as agent technologies mature and are adopted at larger scale.

%%% ROB Shrink
AI agents are moving from experimentation to everyday use. Practitioner accounts show how organizations move from pilots to production, driven by automation potential but constrained by reliability, toolchain limitations, and engineering complexity. Across industries, recurring coordination strategies and reusable patterns show that successful agentic systems rely on explicit architectural structures, including task decomposition and coordination to achieve scalability and maintainability. Differences across domains show that agent-based architectures do not generalize uniformly, as domain constraints, interaction patterns, and technology stacks shape implementation choices and operational behavior. As these systems evolve, understanding how to engineer reliable, controllable, and domain-adapted agent architectures remains a key challenge.

%% file: Sections/bios.tex
% \textcolor{blue}{we need to provide a brief bio for each author}

\begin{IEEEbiography}{Ruoyu Su}{\,}  is a doctoral researcher at the University of Oulu, Finland. He earned his MSc degree in Information Processing Science - Software Engineering from the University of Oulu. His research focuses on the intersection of Large Language Models, Software Architecture, and Empirical Software Engineering. Contact him at ruoyu.su@oulu.fi\vspace*{8pt}
\end{IEEEbiography}

\begin{IEEEbiography}{Matteo Esposito}{\,} is a postdoctoral researcher at the University of Oulu, Finland, with a European Label PhD from the University of Rome ``Tor Vergata''. His research focuses on Generative AI and software architecture, maintenance,  quality, and security. Contact him at matteo.esposito@oulu.fi\vspace*{8pt}
\end{IEEEbiography}

\begin{IEEEbiography}{Roberta Capuano}{\,} is a postdoctoral researcher at University of L'Aquila, Italy, where she earned her Ph.D. in Information and Communication Technology. She is working on quality-driven legacy system modernization using generative AI, with a focus on sustainability–performance trade-offs. Contact her at roberta.capuano@univaq.it.\vspace*{8pt}
\end{IEEEbiography}

\begin{IEEEbiography}{Rafiullah Omar}{\,} is a postdoctoral researcher at the University of L’Aquila, Italy, where he earned his Ph.D. in Information and Communication Technology. He works on Green AI, large language models (LLMs), and LLM-based agentic systems. He can be contacted at rafiullah.omar@univaq.it.

\vspace*{8pt}
\end{IEEEbiography}

\begin{IEEEbiography}{June Sallou}{\,} is an Assistant Professor in the Information Technology Group, at Wageningen University, in the Netherlands. Her current research interests include sustainable software engineering, and Green AI. She received her Ph.D. in Computer Science from the University of Rennes, France. Contact her at june.sallou@wur.nl.\vspace*{8pt}
\end{IEEEbiography}

%\begin{IEEEbiography}{Henry Muccini}{\,} \vspace*{8pt}
%\end{IEEEbiography}
%\usepackage{textcomp} %Add to preamble to use the \textregistered font 

\begin{IEEEbiography}{Henry Muccini}{\,}is a Professor of Software Engineering in the SWEN research group and leader of the FrAmeLab research laboratory at the University of L'Aquila, Italy. His research primarily focuses on Software Architecture, Software Engineering and AI, and Green Software Engineering/Green AI. Contact him at henry.muccini@univaq.it.
\end{IEEEbiography}

\begin{IEEEbiography}{Davide Taibi} {\,} 
is LEGO\textregistered Chair and Professor of Software Engineering at the University of Southern Denmark, 7100 Vejle, Denmark, and a software architecture consultant specializing in cloud-native systems. He can be contacted at davide@taibl.it.
\end{IEEEbiography}